# Financial Management Challenges in Enterprises Employing Remote and Hybrid Workforces

Michał Ćwiąkała[1], Gabriela Wojak[2], Dariusz Baran[3], Ernest Górka[4], Bartłomiej Bartnik[5], Waldemar Gajda[6], Ryszard Ratajski[7]

*Abstract:*

***Purpose:*** *This paper explores financial management challenges in companies with remote and hybrid workforces. It examines how flexible work models affect budgeting, reporting, and transparency. The study aims to identify key factors that influence efficiency in distributed environments.*
***Design/methodology/approach:*** *A quantitative survey was conducted among managers, HR staff, and finance professionals with experience in remote or hybrid work. The questionnaire gathered data on digital tool usage, communication, and financial process effectiveness. Responses were analyzed to assess the impact of work models on financial outcomes.*
***Findings:*** *Remote and hybrid work improve budget control and financial transparency due to digital tools like ERP systems. Forecasting and communication across departments remain major challenges. Respondents also reported lower stress and better work–life balance in flexible work settings.*
***Practical recommendations:*** *Organizations should strengthen digital infrastructure and use advanced analytics for better financial forecasting. Clear communication strategies and employee well-being support are crucial. Remote models require secure systems and accessible documentation to ensure efficiency.*
***Originality/value:*** *This is one of the first empirical studies on financial management in remote and hybrid settings. It offers data-driven insights into budgeting, reporting, and employee outcomes. The findings support strategic decision-making for managers navigating flexible work environments.*

[1]*University College of Professional Education in Wroclaw, Poland,*
*ORCID: 0000-0001-9706-864X, e-mail:* michal.cwiakala@wsk.pl;
[2]*I'M BRAND INSTITUTE Sp. z o.o., ORCID: 0009-0003-2958-365X,*
*e-mail:* g.wojak@imbrandinstitute.com;
[3]*Department of Social Sciences and Computer Science Nowy Sącz School of Business - National Louis University, Poland, ORCID:0009-0006-8697-5459,*
*e-mail:* dkbaran@wsb-nlu.edu.pl;
[4]*The same as in 3, ORCID: 0009-0006-3293-5670, e-mail:* ewgorka@wsb-nlu.edu.pl;

[5] *The same as in 3, ORCID: 0000-0003-2370-3326, e-mail:* bbartnik@wsb-nlu.edu.pl;
[6]*Warsaw School of Management – Higher Education Institution, Poland,*
*ORCID: 0000-0003-0739-4340, e-mail:* waldemar.gajda@wsz-sw.edu.pl;
[7]*Pomeranian Higher School in Starogard Gdanski, Institute of Management, Economics and Logistics, Poland, ORCID: 0000-0002-4863-9714, e-mail:* ryszard.ratajski@twojestudia.pl;

**Paper type:** *Research article.*

**Declaration of interest statement:** *The authors declare that they have no conflict of interest.*

### 1. Introduction

In recent years, remote and hybrid work models have significantly transformed how organizations operate globally. Initially driven by the COVID-19 pandemic, these shifts have reshaped not only daily workflows but also the structures, systems, and processes that underpin financial management (Eurofound, 2020; Kniffin *et al.,* 2021). In this new paradigm, companies are required to adapt their tools, policies, and cost control mechanisms, prompting a fundamental rethinking of traditional financial management practices (Brynjolfsson *et al.,* 2020; Zupok, 2024; Baran *et al.,* 2025).

Although research on remote and hybrid work has grown substantially, the majority of existing studies focus on psychosocial aspects, employee productivity, or the use of digital communication technologies (Felstead and Reuschke, 2020; Wang *et al.,* 2021).

Far less attention has been given to how these work models influence financial decision-making processes, particularly in areas such as budgeting, cost control, financial forecasting, and interdepartmental coordination in a distributed environment. A notable research gap exists in the lack of empirical studies that connect digital infrastructure (such as ERP systems and collaborative platforms) with perceptions of financial efficiency and organizational performance in remote work contexts.

This study is original in its focus on practical challenges and opportunities in financial management within remote and hybrid organizations. Drawing on data collected from employees in operations, HR, and finance departments who work daily in distributed settings, it provides real-world insights into the interplay between technology, communication, and financial control. The research accounts not only for technological enablers (e.g., ERP systems, analytics tools) but also for psychological and organizational dimensions such as stress levels, trust culture, and communication efficiency (Azizzadeh *et al.,* 2022; Islam *et al.,* 2023; Zupok and Dyrka, 2022).

Based on a review of relevant literature and the identified research gap, the following research questions are formulated: What are the key financial management challenges faced by enterprises operating under remote and hybrid work models? How do digital tools and organizational structure influence financial transparency and control in these work environments? What psychological and communication-related factors affect the effectiveness of financial processes in distributed teams?

The structure of this paper includes a literature review on remote work and financial management, a presentation of the research methodology, analysis of empirical findings, and a discussion section with practical recommendations for business leaders and financial managers.

## 2. Literature Review

Across management and labor studies, "remote work" is consistently defined as the performance of job tasks outside the employer's premises, enabled by information and communication technologies (ICT). Polish and international sources converge on these core elements while differing in the emphasis they place on organizational and social consequences. In Polish scholarship, remote work is framed as a non-standard employment form grounded in employee autonomy over time and place (Zając, 2018) and as an organizational model in which dispersed locations are coordinated predominantly through digital means (Bednarski, 2020).

Internationally, definitions highlight the technological substrate (smartphones, laptops, broadband) and the regularity of off-site work, together with challenges for work–life balance and managerial control (Messenger and Gschwind, 2016; Eurofound and ILO, 2017). These positions align with human capital perspectives in Poland that read remote work through the lens of HR strategy and cost optimization (Juchnowicz, 2014).

Hybrid work is commonly conceptualized as a combinatory arrangement integrating on-site and off-site work in a schedule that is either predefined or employee-driven (Deloitte Insights, 2021; Gallup Inc., 2022; Gąciarz, 2022; Skrzek-Lubasińska, 2021; Śliwiński, 2023). Polish authors underscore its organizational embeddedness-coordination, office space design, and employee well-being - within broader digital transformation agendas (Gąciarz, 2022; Skrzek-Lubasińska, 2021; Śliwiński, 2023). Global reports accent flexibility and distributed team performance (Deloitte Insights, 2021; Gallup Inc., 2022).

Within these umbrellas, the literature distinguishes several archetypes: fully remote, fixed hybrid (pre-set on/off-site days), flexible hybrid (employee choice), and rotational models (team rotations); an "anywhere office" variant extends location flexibility across borders, subject to security and compliance constraints (Brinatti, Kowalski and Evans, 2023; Buckley, 2021). Taken together, the models represent an evolutionary response of labor markets and organizations to demands for flexibility, digitalization, and outcome-oriented management (Eurofound and ILO, 2017; Gąciarz, 2022).

National regulation in Poland evolved rapidly under COVID-19 pressures, exposing legal gaps and catalyzing statutory updates that normalized remote and hybrid arrangements. Before the pandemic, adoption and preparedness were limited, 53% of large enterprises lacked readiness - and remote work participation rose to 9.2% in Q2 2020 as policy adapted (Krupska, 2023). Regulatory change is portrayed as necessary but insufficient without employer action on technical standards, organizational rules, and inclusion of vulnerable groups (Krupska, 2023; Białożyt-Wielonek, 2022).

Internationally, heterogeneous rules on taxation, labor, and data protection complicate "work-from-anywhere" policies; many firms (estimated 64-72%) have adopted bespoke frameworks across jurisdictions, incurring additional compliance costs (Cygal, Gilliland, Hannibal, and Stirling, 2021).

Cross-country digitalization gaps further constrain effectiveness, with Poland trailing Western Europe on digital indices (Szewczyk, 2024). The literature also flags data protection as a persistent risk that requires sustained, and often costly, investment in cybersecurity capabilities (Kowalczyk and Mroczko, 2024).

Across sources, the principal organizational benefits are flexibility, access to broader talent pools, and cost reconfiguration. From the employee perspective, time savings are salient, 76.2% indicate remote work saves time - driven by the elimination of commuting and schedule autonomy (Kurnyta, 2023).

Employers gain agility in human resource allocation across locations and specialties (Hassan and Thornley, 2024), while hybrid arrangements reduce logistical overhead (e.g., less inter-meeting travel) and enable more efficient resource planning (Hassan and Thornley, 2024). Firms commonly report lower office-related costs, with 72% citing reduced spending on premises (Kurnyta, 2023).

Remote/hybrid models are also cast as inclusion mechanisms: they lower barriers for individuals with mobility constraints or caregiving responsibilities, can mitigate gender disparities, and expand regional employment opportunities (Williamson, 2022; Kukuczka, 2023). Health and well-being links are noted for hybrid work (more exercise, sleep, better diet), with plausible implications for reduced sickness absence and greater engagement (Hassan and Thornley, 2024).

Benefits are offset by several risks. First, technology and security costs rise - VPNs, MFA, encryption - and must be budgeted and maintained (Kowalczyk and Mroczko, 2024). Second, equity and ergonomics issues emerge when employers under-fund home-office equipment beyond basics, potentially depressing productivity in analytics-intensive roles (Cygal *et al.,* 2021). Third, cybersecurity exposure increases in distributed settings; pandemic-era spikes in attacks have motivated higher spending on controls (Buckley, 2021). Finally, literature emphasizes culture and well-being risks: isolation, blurred boundaries, and burnout, especially absent trust-based leadership and clear norms (Becker, Belkin, Tuskey, and Conroy, 2022; Wysocka, 2021; Wójcik, 2022).

Effectiveness in remote/hybrid work is shown to be multi-causal, spanning technology, culture, skills, security, ergonomics, time management, mental health, and communication transparency (Aziz, 2021; Becker *et al.,* 2022). Critical enablers include:

- robust collaboration and project platforms (e.g., Teams, Slack, Asana/Trello) and finance/analytics systems (QuickBooks, SAP), with the caveat that misuse or under-use diminishes returns (Lipińska, 2021);
- trust-first leadership and autonomy to avoid micromanagement pitfalls (Nowak, 2020);
- sustained digital upskilling, notably ERP/analytics capabilities (Dąbrowski, 2022);
- information security governance (policies, encryption, MFA) to mitigate reputational and financial loss (Kowalski, 2021);
- ergonomic support and well-being programs to protect performance and reduce attrition risks (Wójcik, 2022; Wysocka, 2021).

The transition towards hybrid and online work models has profound implications for employee motivation, engagement, and team performance. Research on motivational factors highlights the growing importance of non-financial incentives such as flexible working hours, additional days off, and supportive work–life balance policies in sustaining workforce motivation in distributed environments (Kasperczuk *et al.,* 2025).

Flexible scheduling in particular emerged as one of the most effective intangible motivators (M = 4.29), underscoring employees' increasing preference for autonomy and control over their working conditions. These findings align with broader trends in the digital workplace, where the ability to integrate professional and personal responsibilities directly enhances job satisfaction and commitment.

Similarly, benefits such as subsidised holidays or reimbursement of commuting costs, which support employees' broader well-being, have been shown to significantly increase engagement levels (Kasperczuk *et al.,* 2025).

Remote/hybrid adoption reshapes cost structures toward IT infrastructure, software licensing, support, and training. Expenditures include laptops/peripherals, secure access (VPN/MFA), software subscriptions (collaboration, ERP, analytics), helpdesk, and periodic hardware refreshes; leasing can smooth capex into opex (Kowalczyk and Mroczko, 2024; Cygal *et al.,* 2021; Kurnyta, 2023).

Licensing strategies (multi-year deals, vendor negotiations) and selective adoption of open-source alternatives are discussed as levers to optimize total cost of ownership (Kowalczyk and Mroczko, 2024; Maczuga and Łais, 2024). While initial outlays are non-trivial, studies argue that process optimization and space savings can offset investments over time (Hassan and Thornley, 2024).

Finance functions increasingly rely on ERP-enabled reporting and analytics to sustain transparency and speed in distributed settings; this shift accelerated during COVID-19 (Bier, Raza, Caram, and Bromberg, 2021; Marecik, 2024). Cloud architectures provide elasticity and real-time data access but require encryption and access-control regimes; migration entails upfront adaptation and training costs (Siemek, 2023; Przybylik, 2024). Automation reduces manual errors and supports predictive budgeting and cost forecasting (Hassan and Thornley, 2024).

The budgeting literature emphasizes continuous monitoring (ERP dashboards, real-time analytics), flexible reallocation, and explicit appraisal of IT ROI and TCO to reflect the dual burden of office and home-workspace costs in hybrid models (Bier *et al.*, 2021; Kowalczyk and Mroczko, 2024). Firms dynamically adjust variable pay and benefits to the chosen work model - about 32% modify pay structures accordingly -while guarding equity and motivation (Cygal *et al.*, 2021).

Performance is measured through ROI (e.g., conferencing, project management, digital communications) and KPI systems; about 24.2% of employers monitor remote productivity via KPIs (Borggreven, 2020; Pokojski and Lipowski, 2022). Evidence also points to non-financial outcomes - lower emissions through reduced commuting, and "soft costs" from disengagement or stress - both of which materially affect the business case (Borggreven, 2020; Flores, 2019).

The reviewed literature portrays remote and hybrid work as strategic configurations rather than ad-hoc accommodations: they re-allocate costs toward digital assets and skills, broaden labor market reach, and enable outcome-centric management (Zupok, 2018; Zupok, 2015; Zupok, 2009). Effectiveness hinges on fit among regulation, technology, culture, skills, and financial controls. Overall, the literature supports the view that remote and hybrid work can enhance organizational efficiency and inclusivity if supported by intentional investments in secure digital infrastructure, skills, and data-driven financial management, under clear regulatory and governance frameworks.

## 3. Research Methodology and Case Description

The conducted research was exploratory in nature, and its primary objective was to identify the factors determining the effectiveness of team management in the new organizational realities, as well as to define the challenges and opportunities arising from the transformation of the work model. This issue stems from the growing importance of remote and hybrid work, which have become a permanent component of contemporary enterprises' operations in the post-pandemic environment. Understanding how organizations can effectively manage dispersed teams is crucial for strengthening their competitiveness, enhancing productivity, and ensuring financial stability.

The adopted research methodology is grounded in a quantitative approach, enabling an analysis of the examined phenomenon from the perspective of a broad group of respondents. This allowed for capturing both general trends and subjective opinions regarding managerial practices, tools employed, and barriers hindering effective collaboration in remote and hybrid work environments. The study encompassed managers, HR specialists, and employees from operational and financial departments who are routinely involved in team management processes within a dispersed work setting.

To collect the data, a questionnaire survey was employed, consisting of three sections. The introductory (address) section provided information about the purpose of the study and assurances regarding the anonymity and confidentiality of responses. The main section included questions related to the communication and management tools used, the effectiveness of collaboration, team productivity, and the barriers encountered. The demographic section contained questions concerning respondents' demographic and professional characteristics, such as position or industry.

In the analysis, two groups of variables were taken into account. The dependent variables included: team work efficiency, communication effectiveness, level of engagement, project coordination quality, and employee stress level. The independent variables, on the other hand, were: work model (remote, hybrid, onsite), digital tools used, and industry.

The adopted methodology made it possible to obtain a broad picture of the phenomenon under study and to conduct an in-depth analysis of managerial practices employed in organizations operating under remote and hybrid work models. The research findings provided valuable insights into the effectiveness of various management strategies and identified factors that support enhancing the effectiveness of collaboration within dispersed teams.

### 4. Research Results

The results of empirical research obtained from the survey are presented below. They have been subjected to detailed analysis, which has enabled a comprehensive assessment of the impact of management strategies and tools on the functioning of teams working in remote and hybrid models. In the first control question, concerning experience of working remotely or in a hybrid mode, all respondents answered in the affirmative. This means that the further results come from the relevant group of respondents who have direct experience of working in such conditions.

**Table 1.** *Distribution of respondents' answers to the question about tools used during remote work*

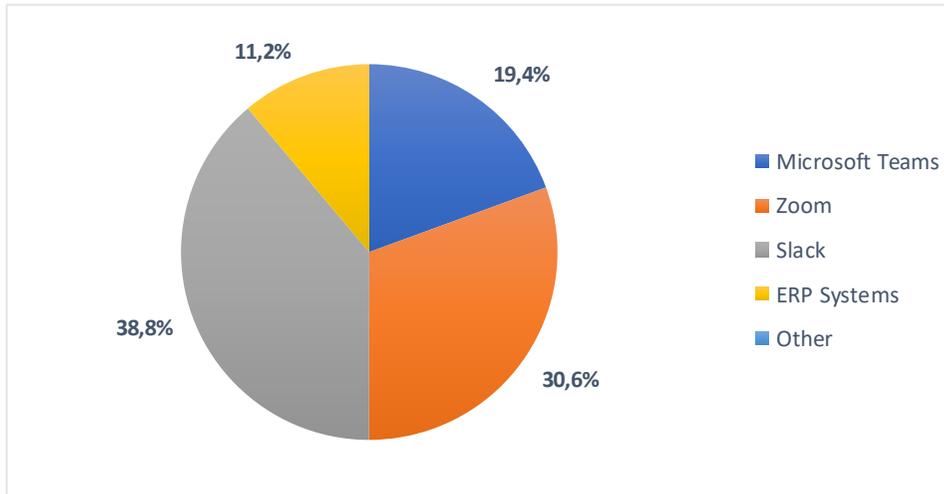

**Source:** *Own elaboration.*

The pie chart above shows the distribution of responses to the question about the most frequently used tools in everyday remote or hybrid work. Analysis of the responses allows us to identify the dominant platforms and systems in the remote work environment of the surveyed group. The responses are distributed as follows:

- Slack: 38.8% of respondents indicated this tool as the most frequently used.
- Zoom: 30.6% of respondents declared that they most frequently use the Zoom platform.
- Microsoft Teams: 19.4% of respondents listed Microsoft Teams as the most frequently used tool.
- ERP systems (e.g., SAP, Comarch): 11.2% of respondents indicated ERP systems as the ones they use most often.

The bar chart shows the distribution of responses to the question about the impact of remote or hybrid working models on the effectiveness of financial management in respondents' organizations. The rating scale ranged from "definitely no impact" to "definitely an impact."

Analysis of the data obtained indicates that the majority of respondents considered that the remote or hybrid work model has a significant impact on financial management – 51.4% of respondents chose the answer "definitely has an impact" and 23.6% chose "rather has an impact." A neutral position was taken by 4.2% of respondents. In turn, 13.9% of respondents believed that this model does not really affect financial management, and 6.9% said that it definitely has no impact.

***Table 2.*** *The impact of remote or hybrid working models on financial management efficiency*

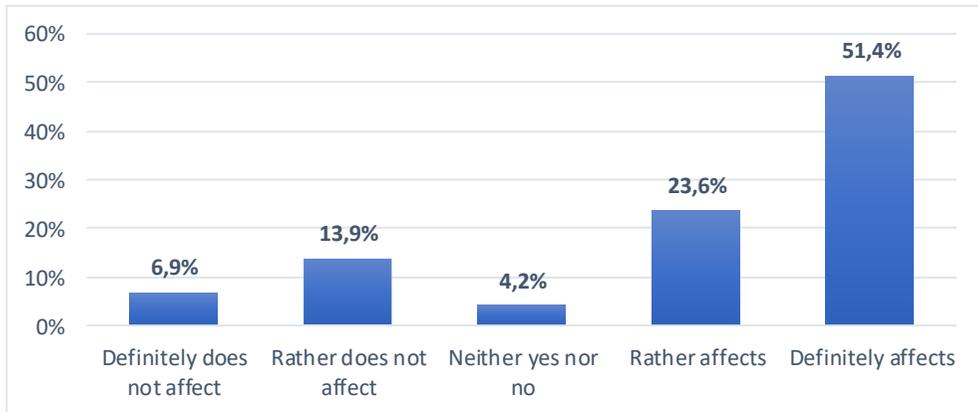

***Source:*** *Own elaboration.*

***Table 3.*** *Assessment of the transparency of financial processes*

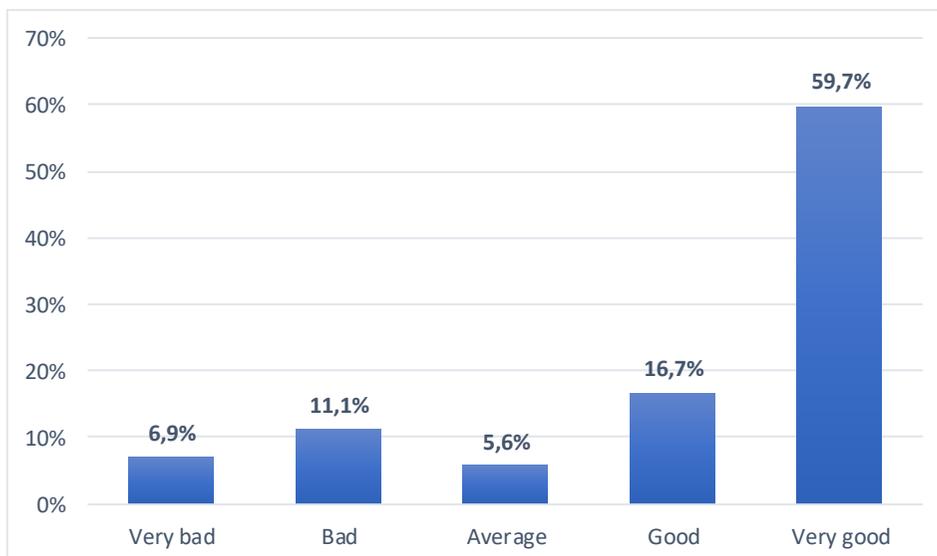

***Source:*** *Own elaboration.*

The bar chart shows the distribution of responses to the question concerning the assessment of the transparency of financial processes, such as invoicing, settlements, and expense approvals, in a distributed work environment. The response scale ranged from "very poor" to "very good." Analysis of the results shows that most respondents positively assess the transparency of financial processes in a remote or hybrid work environment – 59.7% rated it "very good" and 16.7% "good." A neutral rating "average" was given by 5.6% of respondents, while 11.1% rated it as 'poor' and 6.9% as "very poor."

The transition to a distributed work model has not worsened, and may even have improved, the transparency of key financial processes, which may be due to the implementation of appropriate digital tools, document circulation systems, and procedures that facilitate access to information and remote tracking of financial processes.

**Table 4.** *Assessment of control over budget implementation in a remote/hybrid working model*

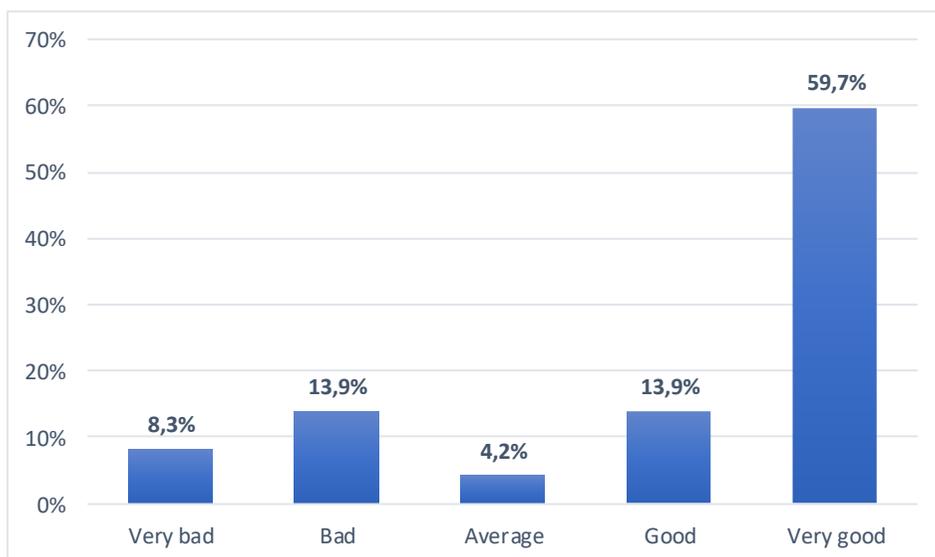

*Source: Own elaboration.*

The bar chart below shows the distribution of ratings for budget control in remote or hybrid working models. The response scale ranges from 1 (very poor) to 5 (very good) and shows the extent to which organizations are able to monitor and manage their budgets in such a working environment. The results show that 8.3% of respondents rated budget control as very poor, 13.9% as poor, 4.2% as average, 13.9% as good, and as many as 59.7% as very good.

This distribution of responses suggests that remote or hybrid work not only did not hinder budget control, but in many cases may have even improved it. This may be the result of the implementation of modern tools for online expenditure monitoring, broader digitization of financial processes, and greater employee responsibility for the implementation of project or departmental budgets.

At the same time, the noticeable percentage of negative assessments shows that for some organizations, financial control in a remote or hybrid model still poses a challenge.

*Table 5. Difficulties in financial management during remote/hybrid work*

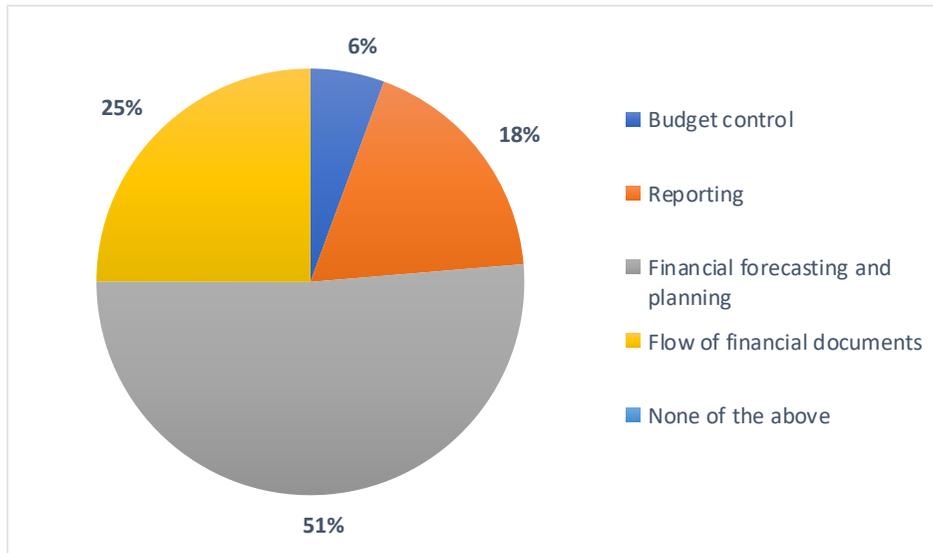

*Source: Own elaboration.*

The pie chart below presents the distribution of responses to the question about the aspects of financial management that are most difficult in remote or hybrid working conditions. The analysis allows us to identify the key challenges that organizations face in flexible working models.

The results show that 5.6% of respondents indicated difficulties related to budget control, 18.1% have problems with reporting, while the biggest challenge is financial forecasting and planning, which was highlighted by as many as 51.4% of respondents. For 25% of respondents, the flow of financial documents is also problematic, while none of the respondents considered any of the above aspects to be difficult.

The next bar chart shows the distribution of responses regarding the extent to which companies incur additional costs related to the implementation of remote or hybrid work. The rating scale ranges from 1 (very high costs) to 5 (no costs) and allows for an assessment of the financial burden on organizations resulting from flexible working models. The results show that none of the companies reported very high costs, 5.6% of respondents rated them as high, 4.2% found it difficult to determine, 25% indicated low costs, while the vast majority – 65.3% – declared that their organization did not incur any additional expenses in this regard.

***Table 6.*** *Assessment of costs incurred by the employer*

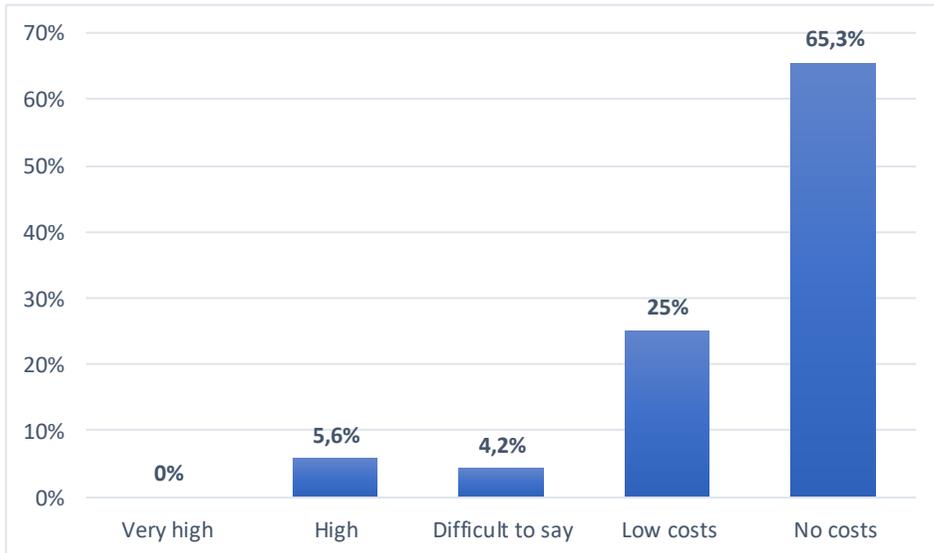

**Source:** *Own elaboration.*

***Table 7.*** *Assessment of communication deterioration in a dispersed team*

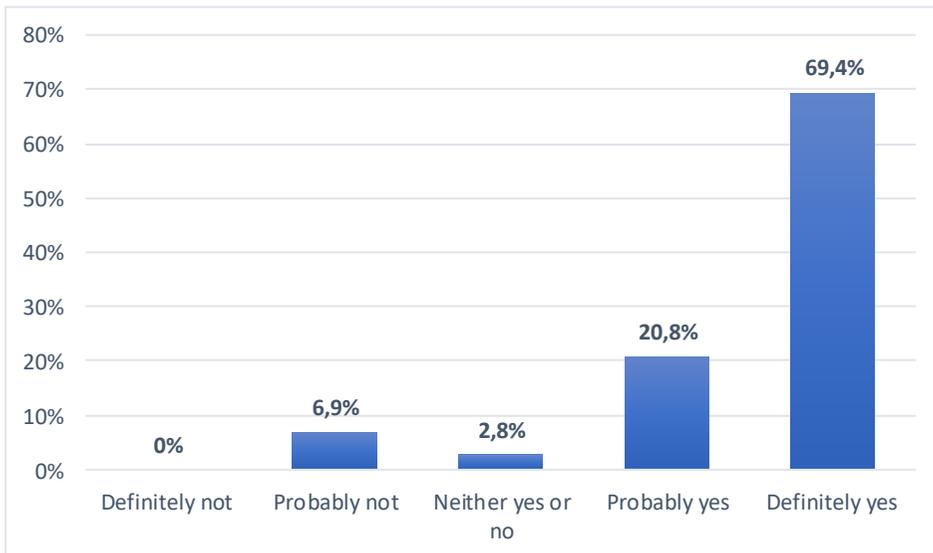

**Source:** *Own elaboration.*

The chart shows the distribution of responses regarding the impact of remote and hybrid work on the quality of communication within a team. The rating scale ranges from strongly disagree to strongly agree and allows us to assess the extent to which respondents perceive a deterioration in communication under flexible working models. The results show that 6.9% of respondents believe that communication has not deteriorated, 2.8% assess it neutrally, while as many as 20.8% indicate that it has deteriorated somewhat, and 69.4% – that it has definitely deteriorated.

Such a clear predominance of assessments indicating a deterioration in the quality of communication suggests that this is one of the key challenges faced by teams working remotely or in a hybrid model. The results highlight the need to implement more effective communication tools and strategies that could support collaboration in a dispersed work environment.

*Table 8.* Challenging areas of communication in distributed teams

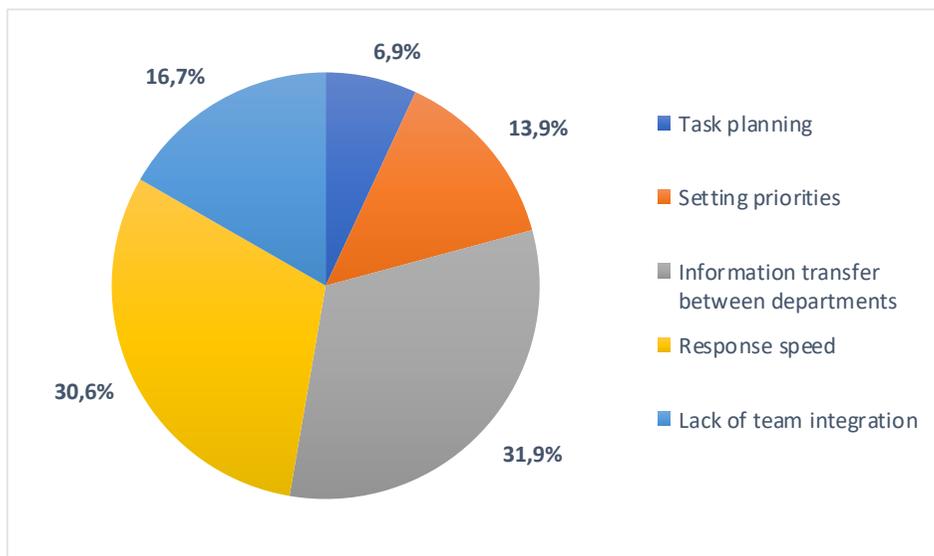

**Source:** *Own elaboration.*

The eighth chart above shows the distribution of responses regarding the areas of communication that pose the greatest challenge in the context of remote or hybrid work. Analysis of the results allows us to identify the key difficulties faced by teams operating in flexible work models.

The data shows that 6.9% of respondents indicated task planning as an area of difficulty, 13.9% have problems with setting priorities, 31.9% considered the transfer of information between departments to be the biggest challenge, and 30.6% pointed to the speed of response. In addition, 16.7% of respondents indicated a lack of team integration as a significant obstacle. The ninth bar chart shows the assessment of stress levels in teams working remotely or in a hybrid model. The response scale ranges from very high stress to no stress and allows us to determine how flexible forms of work affect employee well-being.

*Table 9.* Assessment of stress levels in dispersed teams

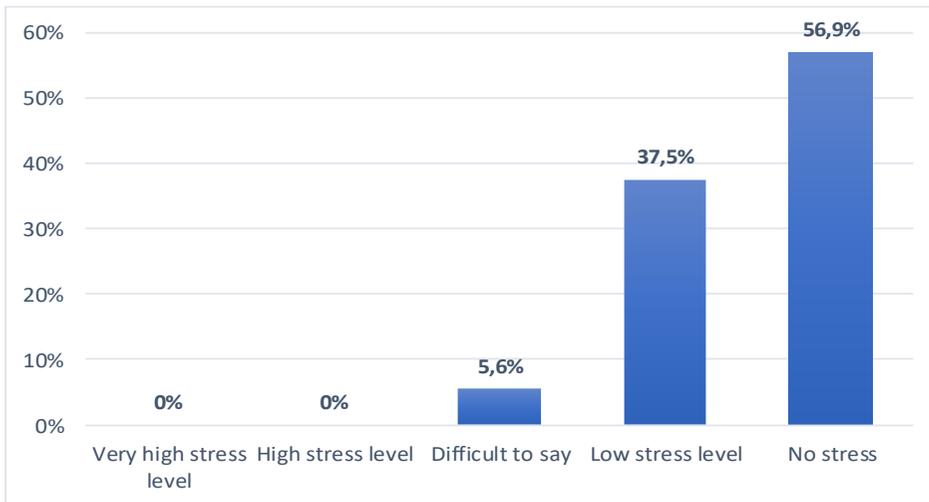

**Source:** *Own elaboration.*

The results show that stress levels are most often rated as low or completely imperceptible in total 94,4% . This may be due to greater flexibility in the organization of working time, greater autonomy, no need for daily commuting, or the ability to more easily combine professional responsibilities with private life. An additional factor contributing to lower stress levels may also be the possibility of creating a more comfortable working environment at home.

*Table 10.* Main sources of stress

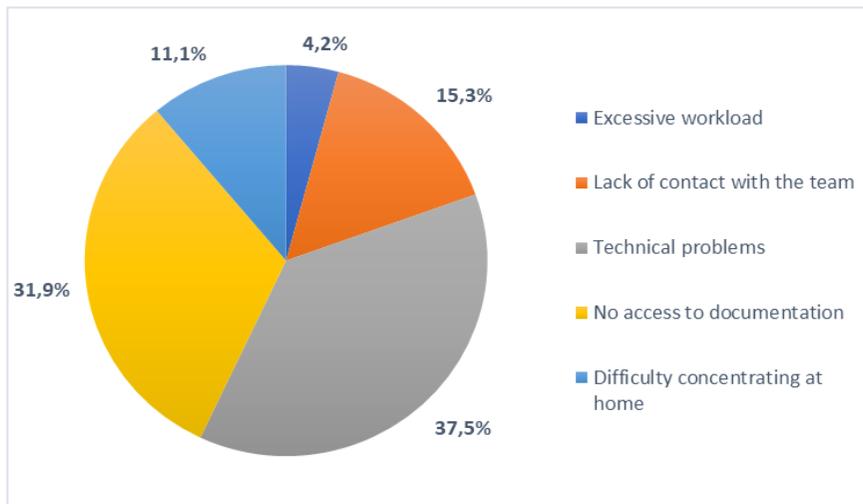

**Source:** *Own elaboration.*

The pie chart shows the main sources of stress among people working remotely or in a hybrid model. The analysis indicates that the most frequently cited cause of stress is technical problems 37.5%, followed closely by lack of access to necessary documentation 31.9%. Limited contact with the team is also a significant factor 15.3%, followed by difficulties concentrating while working from home 11.1%. The least frequently cited source of stress was an excessive workload 4.2%.

These results highlight the importance of reliable IT infrastructure, efficient access to documents, and effective team communication in a distributed work environment. They also draw attention to the need to create conditions conducive to concentration and a balance between professional and domestic responsibilities.

*Table 11. Assessing the impact of distributed work on work-life balance*

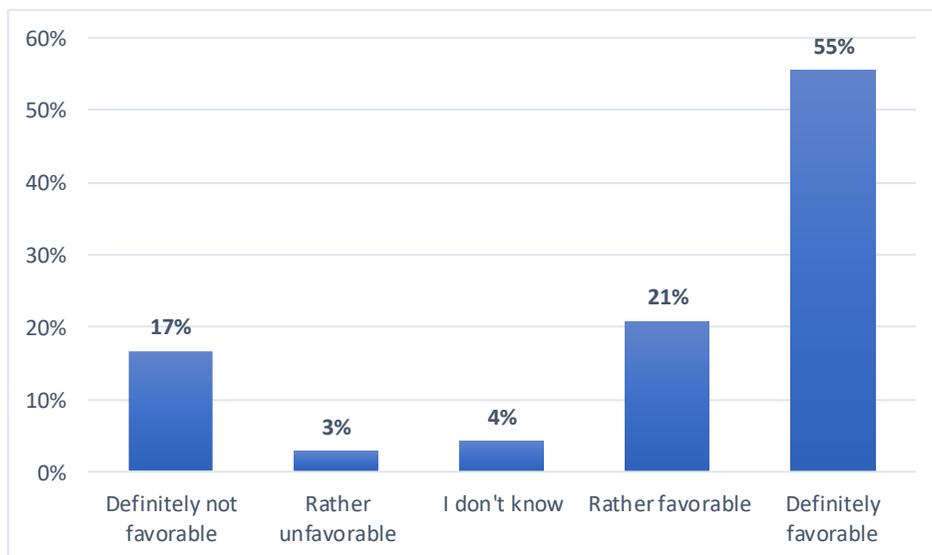

**Source:** *Own elaboration.*

The next bar chart shows the results of the assessment of the impact of remote and hybrid work on work-life balance. The response scale ranges from definitely not favorable to definitely favorable and allows us to determine how respondents perceive the impact of flexible working models on work-life balance.

The analysis of the results shows a very positive assessment – more than half of the respondents 55% believe that remote or hybrid work definitely promotes a better work-life balance, and another 21% assess its impact as rather positive. In total, as many as 76% of respondents see benefits in terms of work-life balance. At the same time, 17% of respondents assess the impact of remote work as definitely negative, and 3% as rather negative, which means that almost one in five people 20% see challenges in it. A small percentage 4% have no opinion on this issue.

These results show that although flexible working models are mostly seen as beneficial, for some employees they can make it hard to keep a good work-life balance. So, organizations should take into account the different needs of their teams and do stuff to support the work-life balance of all employees.

*Table 12. Respondents' place of residence*

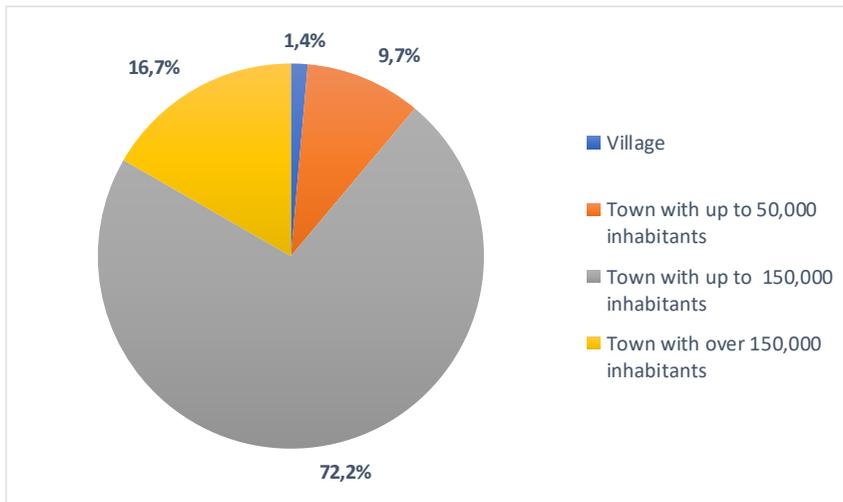

*Source: Own elaboration.*

The vast majority of respondents 72.2% live in cities with up to 150,000 inhabitants. The second largest group 16.7% comes from cities with more than 150,000 inhabitants. A significantly smaller percentage of respondents 9.7% live in cities with up to 50,000 inhabitants. The percentage of respondents living in rural areas is very low.

*Table 13. Respondents' industry*

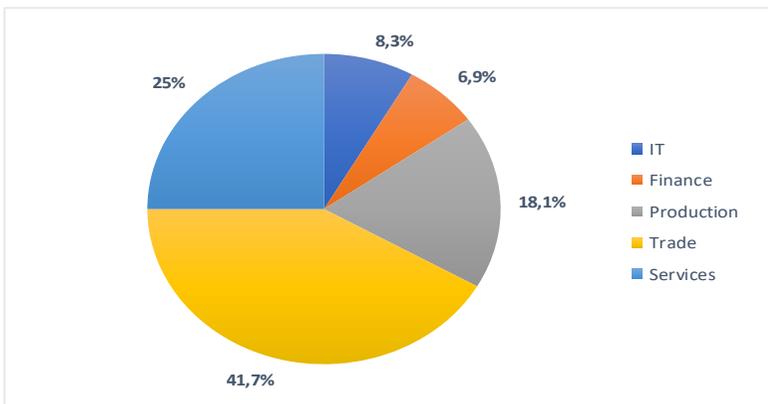

*Source: Own elaboration.*

Analysis of the chart shows that the largest group of respondents 41.7% are people working in trade. The next largest groups are people working in services 25% and manufacturing 18.1%. The smallest percentage of respondents represent the IT 8.3% and finance 6.9% industries.

## 5. Conclusions and Future Research Implications

The conducted research provides comprehensive evidence of how remote and hybrid work models transform key dimensions of organizational functioning, including financial management, communication effectiveness, employee well-being, and work–life balance. The findings clearly indicate that flexible forms of work are no longer temporary solutions but have become integral components of organizational strategies, reshaping managerial practices and operational processes.

From a financial perspective, remote and hybrid work significantly influence budgeting, cost control, and reporting processes. Most organizations report improvements in budget monitoring and transparency of financial procedures, which can be attributed to the digitalization of workflows and the implementation of ERP systems.

However, financial forecasting and long-term planning remain major challenges, consistent with the findings of Hassan and Thornley (2024), who argue that distributed environments require more advanced analytical tools and adaptive financial governance. Importantly, most respondents did not report significant additional costs, indicating the cost-effectiveness of flexible work arrangements over time.

Communication emerged as one of the most critical problem areas. The majority of respondents reported a deterioration in communication quality, particularly in interdepartmental information flow, response speed, and team integration. These results confirm previous literature (Becker *et al.,* 2022; Wysocka, 2021) emphasizing that communication effectiveness in distributed teams is not only a technical challenge but also a cultural one, requiring deliberate strategies, shared digital platforms, and trust-based leadership.

Employee well-being outcomes present a more optimistic picture. Most respondents reported low stress levels, likely due to increased autonomy, time flexibility, and reduced commuting. Nonetheless, technical issues and limited access to documentation were identified as key stressors, underscoring the importance of reliable IT infrastructure and efficient knowledge management systems.

Furthermore, the results demonstrate a strong positive impact of remote and hybrid work on work–life balance, which aligns with recent findings by Kasperczuk *et al.* (2025) highlighting the motivational value of flexible scheduling and non-financial incentives.

Based on the findings, several practical recommendations can be proposed:

- ➢ Investment in digital infrastructure, including enhanced ERP, analytics, and collaboration tools are essential to improve financial forecasting and reporting accuracy.
- ➢ Strategic communication initiatives, such as fostering cross-departmental collaboration, implementing structured feedback channels, and investing in communication training can mitigate communication breakdowns.
- ➢ Support for employee well-being, as an example, organizations should strengthen IT support, ensure secure and easy access to documentation, and design initiatives that promote social interaction and inclusion in distributed settings.

Despite its contributions, the study has several limitations. First, the survey was limited to a specific group of respondents, which may affect the generalizability of the results. Second, the reliance on self-reported data introduces the possibility of response bias. Finally, the cross-sectional design does not capture the dynamic evolution of remote and hybrid work practices over time.

Future research could address these limitations by conducting longitudinal studies and expanding the analysis to different sectors and organizational contexts. It would also be valuable to explore the interplay between remote work practices and key outcomes such as innovation, retention, and organizational culture.

As flexible work models continue to evolve, such studies could provide deeper insights into how digital transformation, leadership, and workforce strategies can be aligned to maximize organizational performance and employee satisfaction in distributed environments.